# Acoustic-Net: A Novel Neural Network for Sound Localization and Quantification

*Guanxing Zhou¹, Hao Liang¹, Xinghao Ding¹, Yue Huang¹, Xiaotong Tu¹,\* and Saqlain Abbas²*

¹ School of Informatics, Xiamen University, Xiamen 361005, China
²Department of Mechanical Engineering, University of Engineering
and Technology (UET) Lahore (Narowal campus), Pakistan
\* Corresponding author: Xiaotong Tu, xttu@xmu.edu.cn

**Abstract:** Acoustic source localization has been applied in different fields, such as aeronautics and ocean science, generally using multiple microphones array data to reconstruct the source location. However, the model-based beamforming methods fail to achieve the high-resolution of conventional beamforming maps. Deep neural networks are also appropriate to locate the sound source, but in general, these methods with complex network structures are hard to be recognized by hardware. In this paper, a novel neural network, termed the Acoustic-Net, is proposed to locate and quantify the sound source simply using the original signals. The experiments demonstrate that the proposed method significantly improves the accuracy of sound source prediction and the computing speed, which may generalize well to real data. The code and trained models are available at https://github.com/JoaquinChou/Acoustic-Net.

## 1. Introduction

Sound source location has been mostly used in the fields of marine, video surveillance and aeronautics. In recent years, beamforming techniques have been proven to be a promising tool for acoustic source localization and quantification. Specifically, given a set of acoustic signals collected by a microphone array, the sound waves have different phases due to different propagation distances. Based on the phase change and amplitude information of the original sound signals recorded by microphones, conventional model-based methods and data-driven deep learning methods have been developed to estimate the sound source location and sound intensity.

Most of the conventional model-based beamforming methods are striving to detect the sound source by improving spatial resolution and reconstructing the true source distribution [1]. The sound source image reconstruction techniques mainly include time-domain and frequency-domain solutions. In time-domain methods, the delay-and-sum (DAS) [2] algorithm is the most commonly used to reconstruct the sound source, but they are limited by the high computational cost. Therefore, the DAS implemented in the frequency-domain is developed with a fast running time. However, it still suffers from limited resolution, which generates a low accurate location result. To sufficiently improve the resolution of sound source localization, Brooks and Humphreys assumed that the sound sources images can be expressed as the linear combination of point spread functions, and proposed the deconvolution approach for the mapping of acoustic sources (DAMAS) principle for incoherent sound sources [3]. Later, they extended the DAMAS to the coherent sound sources and proposed the DAMAS-coherent (DAMAS-C) algorithm [4] with the huge computational burden. Although traditional frequency-domain algorithms can achieve high-resolution images, the applications of these methods still suffer from high computation complexity and need to adjust the grid resolution manually.

Recently, data-driven deep learning methods have been applied to solve acoustic source localization problems. Compared to the model-based methods, deep learning can reconstruct the higher-resolution sound source image with a strong feature extraction capability. In addition, once the network structure is fixed, the process of sound source estimation is found to be faster than the frequency-domain methods. However, the methods (DNN [5], CNN [6], ResNet [7]) with complex network structures are difficult to be integrated into hardware which has limited memory utilization and storage resources.

To overcome the above-mentioned problems, an efficient neural network, the Acoustic-Net, is proposed in the current study, to yield the acoustic localization and quantification as the feature regression problems instead of reconstructing high-resolution images. The Acoustic-Net implicitly models the training amplitude features with shallow one-dimensional (1D) convolution and the training phase change with the RepVGG [8]. Specifically, the main contributions are listed below:
- A novel and comprehensive beamforming dataset is created.
- The proposed method makes the sound source localization and quantification as feature regression tasks, which does not need to obtain high-resolution sound source images by reconstructing sound source distribution.

## 2. Methods

In this section, sound source localization and sound pressure prediction are solved by RepVGG-B0 [8] using the simulation data from 56 spiral array microphones. The beamforming dataset information, the input data preprocessing and the structure of the Acoustic-Net are provided in Section 2.1, Section 2.2 and Section 2.3, respectively.

### 2.1 Beamforming dataset information

The random white noise time signals are simulated by the acoustic software Acoular [9]. The spiral array of 56 microphones is shown in Fig. 1.

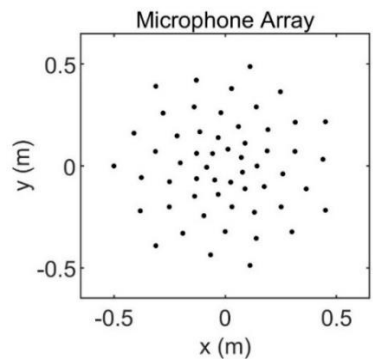

Figure 1: The distribution of 56 spiral array microphones.

Firstly, a novel and comprehensive dataset[1] is introduced for the single sound source localization and sound pressure prediction [10]. The properties of the Beamforming Dataset are summarized in Table 1.

The Beamforming Dataset contains 4200 samples with 2400 points for training, 800 points for validation and 1000 points for testing. Each random white noise signal is collected by 56 spiral array microphones with a sampling frequency of 51200 Hz. An example of the signal sampled by one microphone is given in Fig. 2.

---

[1] The entire dataset is publicly available for download at https://drive.google.com/file/d/1wPeOIcgcrq52-LQXwKE1-VQCMUrIBZuw/view?usp=sharing.

Table 1: Properties of the Beamforming Dataset.

| Experimental setup | Simulated data |
|---|---|
| Microphone array | 56 spiral sensors, as shown in Fig.1 |
| Scanning grid | x, y ∈ [−1.5,1.5] (m) |
| Sampling rate | 51200 Hz |
| Sampling duration | 1 s |
| Source distribution | Uniform |
| Signals | The uncorrelated white noise signal |
| Root mean square (RMS) amplitude of source signal | In the range 0 to 1 (for point source: in 1 m distance) |
| The distance between the sound source point and the microphone array | 2.5 m |

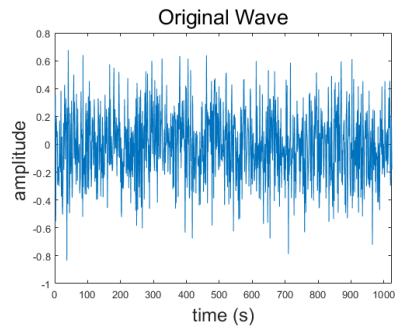

Figure 2: The waveform of the sound signal.

## 2.2 Input Data Preprocessing

To facilitate the training of Acoustic-Net, each signal is transformed into a time-frequency spectrum grey image as presented by Fig. 3. In this work, as the STFT results contain rich time-frequency information, the signals collected by 56 spiral array microphones are transformed into time-frequency distribution exploiting short-time Fourier transform (STFT) [11]. To remove unnecessary redundancy, 56 STFT results are converted to grayscale as the input of the network. The STFT results $y = [y_1, y_2, \cdots, y_N]$ at $N = 56$ are formed by a short-time Fourier transform of the raw sound data $x = [x_1, x_2, \ldots, x_N]$ according to Eq. (1).

$$y_i = \sum_{n=-\infty}^{+\infty} x_i(n)W(n-mR)e^{-j\omega_i n}, \qquad i = 1, 2, \cdots, N, \tag{1}$$

where $W(n)$ represents the length of the Hamming function, whereas $R$ indicates the hop size between the neighboring windows.

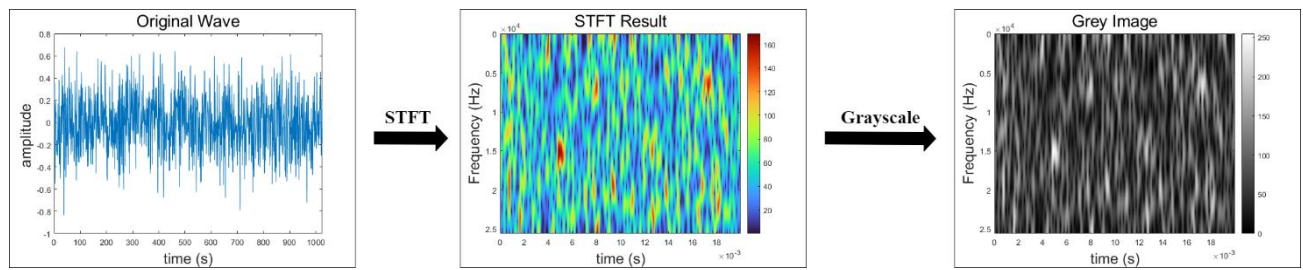

Figure 3: Preprocessing of the sound data.

## 2.3 Description of Acoustic-Net

The Acoustic-Net is designed to extract the features of amplitude change and phase difference by using raw sound data. Since the tasks of sound source localization and sound pressure prediction are not valuable, the Acoustic-Net uses RepVGG-B0 [8] to observe the phase relationship among the 56 STFT grey images. To highlight the amplitude relationship, Acoustic-Net introduces multi-task learning (MTL) to process the raw sound data. A schematic overview of the entire sound source location and sound source pressure prediction is given in Fig. 4.

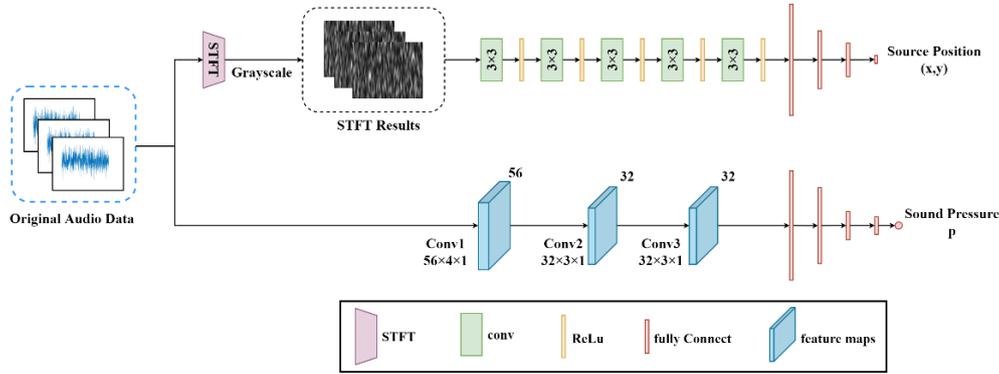

Figure 4: Schematic overview of Acoustic-Net.

### 2.3.1 RepVGG-B0

The re-parameterization VGG (RepVGG) [12] is a simple system with only one single type of operator: 3×3 conv followed by ReLU. Compared with the VGG [13], ResNet and DenseNet [14], the RepVGG can save more computing resources and achieve fast computing speed, with fewer operators and a single convolutional layer when implemented to a hardware device.

The RepVGG has a variety of specific network structures. To extract the phase features with apposite network parameters, the RepVGG-B0 is chosen as the backbone. During the training procedure, RepVGG-B0 uses identity and 1×1 branches as shown in Fig. 4(a). After training, the identity and 1×1 branches can be removed by structural re-parameterization. During the inference time, RepVGG-B0 can convert a trained block into a single 3×3 conv layer as given in Fig. 4(b). Only 3×3 conv is suitable to be optimized by some computing libraries like NVIDIA cuDNN and Intel MKL on GPU and CPU.

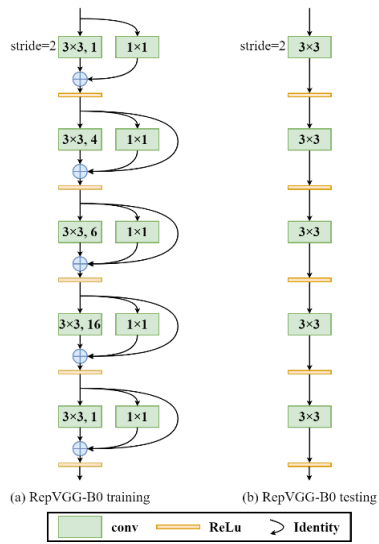

Figure 5: Diagram of RepVGG-B0 architecture: 5 stages, and the layers of each stage are [1, 4, 6, 16, 1].

*2.3.2   Multi-Task Learning*

The RepVGG-B0 has advantages in model simplification and extracting the phase information. However, advanced amplitude information should be extracted from the raw sound data. Thus, the Acoustic-Net is re-implemented by using multi-task learning.

Multi-task learning [15] has been successfully applied in natural language processing and computer vision. The MTL can bias the model to pay more attention to the representation of different features of the original data, so that the model can be better generalized to new tasks. The process diagram of proposed MTL network is shown in Fig. 6. This network structure contains 3 layers of 1D convolution and 4 fully connection layers, as shown in Fig. 4.

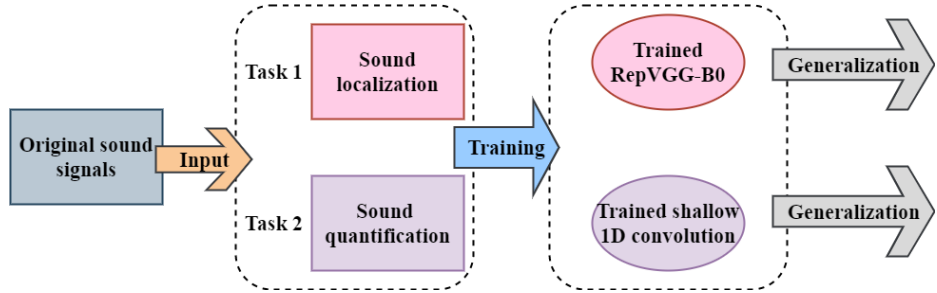

Figure 6: The process diagram of MTL. The task 1 is sound localization, while task 2 is sound quantification.

## 3.   Experiments

In this section, the model training, evaluation metric and experiment results are presented. To demonstrate the effectiveness of Acoustic-Net for sound source localization and quantification, the experiments are conducted on the beamforming dataset, which contains 2400 training samples, 800 validation samples and 1000 samples for test. The results are compared with state-of-the-art sound localization methods.

### 3.1   Model Training

Given the training beamforming dataset $\{Y^r := [x, y, p]^T\}_{r=1}^R$, where $R = 2400$. The vector $Y^r$ contains the acoustic position $\{X^r := [x, y]^T\}_{r=1}^R$ and its strength $\{P^r := [p]^T\}_{r=1}^R$. Let $\hat{X}^r := [\hat{x}, \hat{y}]^T$ denote the predicted sample corresponding to each training sample $X^r$. The location distance loss is minimized on each sample:

$$L_D(X^r, \hat{X}^r) = \|X^r - \hat{X}^r\|_2^2, \qquad (2)$$

where the ℓ2-norm is used to measure the distance between predicted location values and ground truth. $p$ is transformed from a linear scale to a decibel scale according to Eq. (3).

$$p_{SPL} = 20 log_{10}(\frac{p}{p_{ref}}). \qquad (3)$$

Generally, $p_{ref} = 20\mu Pa$ indicates the reference value of sound pressure in the air, and the unit of $p_{SPL}$ is dB. Let $\hat{P}^r_{SPL} := [\hat{p}_{SPL}]^T$ denote the predicted sound pressure level (SPL). During training, the ℓ1-norm loss is minimized as SPL loss:

$$L_{SPL}(P^r_{SPL}, \hat{P}^r_{SPL}) = \|P^r_{SPL} - \hat{P}^r_{SPL}\|. \qquad (4)$$

Combining loss functions in Eq. (2) and Eq. (4), the training objective for Acoustic-Net is constructed as:

$$L(\theta_d, \theta_{spl}) = \frac{1}{N}\sum_{r=1}^{N}\left(\alpha L_D(X^r, \hat{X}^r) + L_{SPL}(P_{SPL}^r, \hat{P}_{SPL}^r)\right), \tag{5}$$

where $\alpha$ is a hyper-parameter in training.

Then, our training methods are described in detail as follows. First of all, the Acoustic-Net can be trained by using Pytorch1.1.0 with a minibatch equal to 8 and under the regularization of $L(\theta_d, \theta_{spl})$ for 150 epochs, during which the momentum of the learning rate is 0.9. Afterwards, the network is implemented using the optimizer Adam with an initial learning rate of 0.01. Finally, the proposed model is trained and tested on a platform with Intel Xeon E5-2678 CPU and NVIDIA TITAN X GPU.

### 3.2 Evaluation Metrics

During the inference procedure, to compare the performance of the commonly used frequency-domain algorithms, the location estimation mean-distance-errors (MDEs), sound pressure level mean-absolute-errors (MAEs) and mean-absolute-percentage-errors (MAPEs) are used to evaluate the performance:

$$MDE_X = \frac{1}{M}\sum_{i=1}^{M} \|\hat{X}^m - X^m\|_2, \tag{6}$$

$$MAE_{SPL} = \frac{1}{M}\sum_{i=1}^{M} |\hat{P}_{SPL}^m - P_{SPL}^m|, \tag{7}$$

$$MAPE_X = \frac{100}{M}\sum_{i=1}^{M} \frac{\|\hat{X}^m - X^m\|_2}{\|X^m\|_2}, \tag{8}$$

$$MAPE_{SPL} = \frac{100}{M}\sum_{i=1}^{M} \frac{|\hat{P}_{SPL}^m - P_{SPL}^m|}{|P_{SPL}^m|}, \tag{9}$$

where $M$ denotes the number of the test data, $\hat{X}^m$ the estimated acoustic location, $X^m$ the real acoustic location, whereas $\hat{P}_{SPL}^m$ and $P_{SPL}^m$ denote the predicted sound pressure level and the real sound pressure level, respectively.

### 3.3 Experiment Results

A total of 1000 samples are used for the test. The Acoustic-Net is compared with several common beamforming methods, including the DAS, DAMAS, clean-point-spread-function (CLEAN-PSF) [16], a clean method based on spatial source coherence (CLEAN-SC) [17] and Fourier-based fast iterative shrinkage thresholding algorithm (FFT-FISTA) [18]. The example point (-0.03, -0.21) of ground truth and prediction intuitively are visualized for test data by different methods as shown in Fig. 7. The red box of each method is enlarged in the upper right corner.

After that, for the task of locating and quantifying the sound source, $MDE_X$ and $MAE_{SPL}$ are used to measure the bias of the accuracy. Meanwhile, $MAPE_X$ and $MAPE_{SPL}$ are aimed at evaluating the variance between the prediction and ground truth. The real-time applications are critical to sound source localization and quantification. Thus, the running time of test data is calculated by different algorithms, respectively.

To adjust the SPL loss and the location distance loss, α = 1, 10 and 100 are conducted for the experiment. As shown in Table 2, the proposed Acoustic-Net with α = 10 generally outperforms the compared methods. On the one hand, the bias and variance of the value estimated by our approach are smaller than other algorithms. On the other hand, the Acoustic-Net runs faster than the competitors, which proves that it may meet the requirement of real-time application.

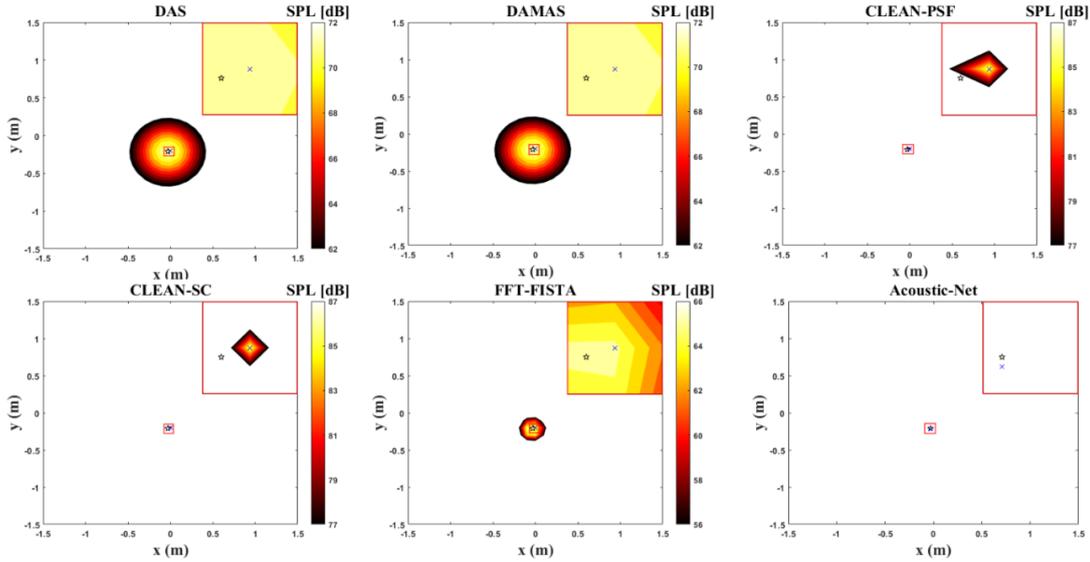

Figure 7: The predicted (×) and the actual (☆) localization of sound source point (-0.03, -0.21). In the red box of each method, the range of x coordinate (m) is [-0.05, 0.05], and the range of y coordinate (m) is [-0.25, -0.15].

Table 2: The MDEs, MAPEs, and MAEs of the frequency-domain methods and Acoustic-Net.

| Methods | $MDE_X(m)$ | $MAE_{SPL}(dB)$ | $MAPE_X$ | $MAPE_{SPL}$ | Time(s) |
|---|---|---|---|---|---|
| DAS | 0.0388 | 1.5583 | 4.6632% | 2.3362% | 0.5030 |
| DAMAS | 0.0388 | 1.5583 | 4.6632% | 2.3362% | 1.3710 |
| CLEAN-PSF | 0.0388 | 13.1364 | 4.6632% | 19.8457% | 0.7094 |
| CLEAN-SC | 0.0388 | 13.1791 | 4.6632% | 19.9102% | 0.6660 |
| FFT-FISTA | 0.0480 | 8.4926 | 5.4585% | 12.8208% | 2.4462 |
| Acoustic-Net$_{\alpha=1}$ | 0.0249 | 0.8601 | 4.5105% | 1.3423% | 0.1544 |
| Acoustic-Net$_{\alpha=100}$ | **0.0061** | 0.7854 | **0.6731%** | 1.1880% | **0.1534** |
| Acoustic-Net$_{\alpha=10}$ | 0.0114 | **0.5641** | 3.6150% | **0.8693%** | **0.1534** |

## 4. Conclusion

In this research, An Acoustic-Net is proposed to locate and quantify the acoustic source without reconstructing images and being restricted by CB map. The acoustic-Net, with RepVGG-B0 and shallow one-dimensional convolution, is used to extract the amplitude and phase characteristics of original sound signals. The proposed method can be easily realized by the hardware devices with fast computing speed and limited memory utilization. The results demonstrate the effectiveness and advantages of the proposed method as compared to the traditional sound localization methods on a beamforming dataset, which is available to investigate the sound source localization and quantification.

## REFERENCES

[1] R. Merino-Martınez, P. Sijtsma, M. Snellen, T. Ahlefeldt, J. Antoni, C. Picard, E. Sarradj, H. Siller, D. G. Dick, and C. Spehr, "A review of acoustic imaging methods using phased microphone arrays," *CEAS Aeronautical Journal*, vol. 10, no. 1, pp. 197–230, 2019.

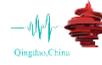

[2] B. Rafaely, "Phase-mode versus delay-and-sum spherical microphone array processing," *IEEE signal processing Letters*, vol. 12, no. 10, pp. 713-716, 2005.

[3] T. F. Brooks, W. M. Humphreys, "A deconvolution approach for the mapping of acoustic sources (DAMAS) determined from phased microphone arrays," *Journal of Sound and Vibration*, vol. 294, no. 4-5, pp. 856-879, 2006.

[4] T. F. Brooks, W. M. Humphreys, "Extension of DAMAS phased array processing for spatial coherence determination (DAMAS-C)," in *12th AIAA/CEAS Aeroacoustics Conference (27th AIAA Aeroacoustics Conference)*, pp. 2654, May 08 – 10, Cambridge, Massachusetts, 2006.

[5] J. Yangzhou, Z. Ma and X. Huang, "A deep neural network approach to acoustic source localization in a shallow water tank experiment," *The Journal of the Acoustical Society of America*, vol. 146, no. 6, pp. 4802-4811, 2019.

[6] W. Liu, Y. Yang, M. Xu, L. Lü, Z. Liu and Y. Shi, "Source localization in the deep ocean using a convolutional neural network," *The Journal of the Acoustical Society of America*, vol. 147, no. 4, pp. EL314-EL319, 2020.

[7] K. He, X. Zhang, S. Ren and J. Sun, "Deep residual learning for image recognition," in *Proceedings of the IEEE conference on computer vision and pattern recognition*, pp. 770–778, 2016.

[8] X. Ding, X. Zhang, N. Ma, J. Han, G. Ding and J. Sun, "RepVGG: Making VGG-style ConvNets Great Again," *arXiv preprint arXiv:2101.03697*, 2021.

[9] E. Sarradj and G. Herold, "A Python framework for microphone array data processing," *Applied Acoustics*, vol. 116, pp. 50–58, 2017.

[10] G. Herold and E. Sarradj, "Performance analysis of microphone array methods," *Journal of Sound and Vibration*, vol. 401, pp. 152–168, 2017.

[11] D. Griffin and J. Lim, "Signal estimation from modified short-time Fourier transform," *IEEE Transactions on Acoustics, Speech, and Signal Processing*, vol. 32, no. 2, pp. 236–243, 1984.

[12] X. Ding, Y. Guo, G. Ding and J. Han, "Acnet: Strengthening the kernel skeletons for powerful cnn via asymmetric convolution blocks," in *Proceedings of the IEEE/CVF International Conference on Computer Vision*, pp. 1911-1920, 2019.

[13] K. Simonyan and A. Zisserman, "Very deep convolutional networks for large-scale image recognition," *arXiv preprint arXiv:1409.1556*, 2014.

[14] G. Huang, Z. Liu, L. Van Der Matten and K. Weinberger, "Densely connected convolutional networks," in *Proceedings of the IEEE conference on computer vision and pattern recognition*, pp. 2261–2269, July 21 – 26, Honolulu, HI, USA, 2017.

[15] S. Ruder, "An overview of multi-task learning in deep neural networks," *arXiv preprint arXiv: 1706.05098*, 2017.

[16] J. A. Högbom, "Aperture synthesis with a non-regular distribution of interferometer baselines," *Astronomy and Astrophysics Supplement Series*, vol. 15, pp. 417-426, 1974.

[17] P. Sijtsma, "CLEAN based on spatial source coherence," *International journal of aeroacoustics*, vol. 6, no. 4, pp. 357-374, 2007.

[18] A. Beck, M. Teboulle, "A fast iterative shrinkage-thresholding algorithm with application to wavelet-based image deblurring," in *IEEE International Conference on Acoustics, Speech and Signal Processing*, pp. 693-696, April 19 – 24, Taipei, Taiwan, 2009.